\documentstyle[12pt,epsf]{article}

\newcommand{\beq}{\begin{equation}}
\newcommand{\eeq}{\end{equation}}

\newcommand{\tra}{{\rm tr}}
\newcommand{\tram}{{\rm Tr}}

\newcommand{\idel}{i \not \! \partial}
\newcommand{\insl}{i \not \! n}
\newcommand{\argu}{- i \theta (\vec{\tau} \cdot \vec{n}) \gamma^5}

\newcommand{\sens}{{\rm sen}}

\newcommand{\qcd}{QCD}
\newcommand{\mitbag}{MIT}
\newcommand{\psir}{\overline{\Psi}}

\title{Cheshire Cat Scenario in a $3+1$ dimensional Hybrid
Chiral Bag}
\author{M.~De~Francia\thanks{e-mail: defranci@dartagnan.fisica.unlp.edu.ar},
H.~Falomir\thanks{e-mail: falomir@dartagnan.fisica.unlp.edu.ar} and
E.~M.~Santangelo\thanks{email: mariel@dartagnan.fisica.unlp.edu.ar}\\
Departamento de F\'{\i}sica \\
Facultad de Ciencias Exactas \\
Universidad Nacional de La Plata \\
C.C. 67 (1900) La Plata - Argentina\thanks{This work was partially supported
by CONICET}}
\begin{document}
\maketitle
\begin{abstract}
The total energy in the two-phase chiral bag model is studied, including the
contribution due to the bag (Casimir energy plus energy of the valence
quarks), as well as the one coming from the Skyrmion in the external sector.

A consistent determination of the parameters of the model and the
renormalization
constants in
the energy is performed.

The total energy shows an approximate independence with the bag radius
(separation limit between the phases), in agreement with the Cheshire Cat
Principle.

{\bf Pacs:} 12.40.Aa

\end{abstract}
hep-ph/9507347
\newpage
\bibliographystyle{unsrt}

The hybrid chiral bag \cite{rhoreport,vento} is an effective model to describe
the
behavior of strongly interacting baryons. In this model, color degrees of
freedom
are confined
to a bounded region and coupled to a bosonic external field (skyrmion) through
boundary
conditions.

These two phase models are intermediate
between two successful descriptions of baryons: bag models \cite{mit,quiral} --
with
\qcd \
degrees of freedom at short distances -- and Skyrme model
\cite{skyrme,witten,adkins}, an
effective (non renormalizable) nonlinear sigma model, useful when the low
energy
properties of
baryons are considered.

An interesting feature of Chiral Bag Models (CBM) is the appearance of the so
called
Cheshire
Cat
Principle (CCP) \cite{rhoreport,nadkarni}, according to which fermionic degrees
of
freedom can be replaced  by bosonic ones in certain regions of space, the
resulting
position of the limit of separation between the two phases having no physical
consequences.

 In $1+1$-dimensions, the Cheshire Cat behavior follows from the bosonization
of
fermionic
fields
\cite{rhoreport}.
In the $3+1$ case, topological quantities, such as the baryonic number, have a
similar behavior \cite{goldstone} but, for non topological ones, the CCP is
expected
to be only
approximately valid.

In what follows, we will study  the energy of a four-dimensional hybrid model
consisting of
quarks and gluons confined to a spherical bag plus a truncated exterior Skyrme
field
in a
hedgehog
configuration. It is our aim to study the dependence of the total energy on the
size
of the
bag, thus testing the Cheshire Cat hypothesis.

\bigskip

In bag models, quarks and gluons are confined to a bounded region that, in our
case,
will be
taken as an static sphere of radius $R$. Adequate boundary conditions are
imposed on
the field
so as to guarantee the vanishing of the
flux of color current to the external sector.

In the \mitbag \ bag model \cite{mit} the boundary conditions for fermions and
gluons
are
\beq
\left. B \Psi \right|_B = \left. \frac12 \left( {\cal I} + \insl \right) \Psi
\right|_B = 0
\label{cc-ferm-mit}
\eeq
\beq
\left. n_\mu F^{\mu \nu} \right|_B = 0 \, .
\label{cc-glu-mit}
\eeq
The Dirac operator for the fermionic field, together with the boundary
condition
(\ref{cc-ferm-mit}) define an
elliptic boundary value problem \cite{fat}. Moreover, (\ref{cc-ferm-mit}) gives
a
vanishing
current at the boundary,
$\left. \psir n_\mu \Psi \right|_B = 0$.

But the axial symmetry, present in \qcd \ with massless quarks, is broken at
the
boundary when
condition (\ref{cc-ferm-mit}) is satisfied. To avoid such unwelcome behavior,
CBM
\cite{quiral} have been proposed, which amount to imposing on the fermionic
field the
boundary condition
\beq
\left. A \Psi \right|_B = \left. \frac12 \left( {\cal I} + \insl e^{\argu}
\right) \Psi \right|_B = 0 \, ,
\label{cc-quiral}
\eeq
where $\tau_i$ are Pauli matrices, and adding an external sector to the bag.

Through this boundary condition, the fermionic field is connected with the
external
field
represented by the ``chiral angle" $\theta (R)$. As it will be shown later, the
external
phase can be described using the Skyrme model.

To obtain the energy in the chiral bag model we will proceed in steps, using
some
results
previously obtained in references \cite{bolsa-quiral,mitdef}.

In the first place, we will introduce the difference between Casimir energies
of
chiral and
\mitbag \ fermionic bags. It corresponds to the zero temperature limit, $T
\rightarrow
0$, of the
results presented
in \cite{bolsa-quiral}. As a second step, we will study the reference vacuum
energy,
i.e., the
\mitbag \
Casimir energy, which is the $T \rightarrow 0$ limit of the cases analized in
\cite{mitdef}
for the fermionic and gauge fields.
An external Skyrme field will then be introduced, so as to complete the two
phase
model (TPM).
The caracteristic parameters in its Lagrangian, together with the
renormalization
constants in
the bag energy, will be determined through physical considerations, thus
obtaining
the total
energy of the TPM.

\bigskip

Determinants
of quotients of elliptic differential operators under different boundary
conditions
can be
expressed as $p$-determinants of quotients of Forman's operators
\cite{forman,pdet,barraza}.
This leads, for a bounded euclidean time, to the study of
differences of free energies of the physical system subject to two different
boundary
conditions. In such a way, in reference \cite{bolsa-quiral} the difference
between
the free
energies of an $SU(2)$ chirally symetric system of massless fermions, confined
to a
spherical
region and
subject to chiral and \mitbag \ boundary
conditions respectively, was calculated.

To construct Forman's operator, which is totally defined by its action over
functions in the
kernel of
$\idel$, a discrete basis of the space of solutions was considered. The
differential
operator
and the
boundary conditions are invariant under the diagonal subgroup of
$SU(2)_{rotation}
\otimes
SU(2)_{isospin}$, and leave
invariant the subspaces caracterized by the quantum numbers $\{k,j,l,m\}$,
associated with
the eigenvalues of
$\left\{ {\bf K}^2, {\bf J}^2, {\bf L}^2, K_z \right\}$,
where ${\bf K}= {\bf L} + {\bf S} + {\bf I}$. In particular, $k$ takes
non-negative
integer
values.

The $T \rightarrow 0$ limit of the above mentioned diference of free energies
leads to
\[
\Delta e_c (\theta)= R \left[ E_c (R,\theta) - E_{c,MIT} \right]
\]
\[
=3
\left\{\frac1{4 \pi} \left[ 4 \pi K_Q \sin^2 \theta + 0.463 \sin^4 \theta +
0.023 \sin^6
\theta
\right] \right.
\]
\[
\left. - \frac1{2 \pi} \sum_{k=1}^{\infty} \int_0^\infty dx \, \left[
2 \nu \log \left( 1 + C_k (x) \sin^2 \theta + D_k (x) \sin^4 \theta -
\Delta (k,x) \right)
\right] \right.
\]
\[
\left.
+\frac{1}{4 \pi}
\left\{
\begin{array}{cr}
\theta^2 & 0<\theta \leq \pi/2 \\
(\pi - \theta)^2  & \pi/2 < \theta < \pi
\end{array}
\right\} \right.
\]
\beq
\left.
- \frac{1}{2 \pi}
\int_0^\infty dx \, \log
\left[
\frac{1 + \frac{\alpha (x)}{x} \frac{-4 + \left[ 2 (1-a^2 (x))^2 +
\frac{\alpha (x)}{x} \right] \cos^2 \theta}
{4 a^2 (x) + (1-a^2 (x))^2 \cos^2 \theta}}
{\left( 1 - \frac{\alpha (x)}{x} \frac1{1+a^2 (x)} \right)^2}
\right] \right\} \, ,
\label{deltaec}
\eeq
where the Casimir energies have been adimensionalized.
The first and second terms in the r.h.s. of (\ref{deltaec}) correspond to the
subspaces $k
\geq 1$. The third and fourth
ones, to the $k=0$ case. We have used the definitions
\[
C_{k}(x) = \frac{-2}{\left[ 4x^2d_{k}^2 (x)+\left( \rho ^2-d_{k}^2 (x)\right)^2
\right]^2}
\times
\]
\[
\left\{
\left( \rho ^2-d_{k}^2 (x) \right) ^2\left[
4x^2d_{k}^2 (x)+
\left( \rho ^2-d_{k}^2 (x) \right) ^2\right]
\right.
\]
\beq
\left.
+\left( \rho
^2-\nu ^2\right) \left[ 4x_n^2d_{k}^2 (x)-\left( \rho ^2-d_{k}^2 (x)\right)^2
\right]
\right\}
\eeq

\bigskip

\beq
D_{k}(x)= \frac{\left[ \left( \rho ^2-d_{k}^2 (x)\right) ^2-\left( \rho
^2-\nu ^2\right) \right] ^2}{\left[ 4x^2d_{k}^2 (x)+\left( \rho
^2-d_{k}^2 (x) \right) ^2\right] ^2}
\eeq

\bigskip

\[
\nu =k+1/2 \qquad \rho =\sqrt{x^2 + \nu^2}
\]
\[
d_k (x) = x \frac d{dx}\ln I_\nu \left( x \right).
\]
\[
a (x)= \coth x  \qquad \alpha (x) = 2 a(x) - \frac1{x} \, ,
\]
where $I_\nu \left( x \right)$ is the modified Bessel function.
In  (\ref{deltaec}), $\Delta (k,x)$ represents the first few
terms\footnote{Note
that, in order to isolate divergences, it is enough to retain the first three
terms
in the asymptotic expansion \cite{bolsa-quiral}. However, in the present
calculation,
we have retained the first six terms in the Debye expansion for computational
convenience.}
in the asymptotic (Debye)
expansion of the other terms in the argument of the logarithm, required to
isolate
the
divergent pieces in the Casimir energy.

In obtaining (\ref{deltaec}), an analytic regularization of non-absolutely
convergent series
has been performed through the introduction of the factor $\rho^{-s}$, for
$\Re{(s)}$ large
enough, and then taking the finite part at $s=0$ \cite{bolsa-quiral}, giving
rise to
the first term in the r.h.s. This procedure leaves an
arbitrary finite part
proportional to $\sin^{2}(\theta)$, which requires the introduction of the
undetermined constant
$K_Q$. This amounts to the introduction, in the Lagrangian of
the external Skyrme field, of the counterterm \cite{zahed}
\beq
\frac{K_0}{16 \pi R} \int_{r=R} d^2x \, \tra \left\{ L_\alpha L_\alpha -
(n_\alpha L_\alpha )^2 \right\} = \frac{K_0}{R} \sens^2 \theta,
\eeq
where $L_\alpha = U^{\dag} \partial_\alpha U = e^{-i \theta (\vec{\tau}
\cdot \check{r} )} \partial_\alpha e^{i \theta (\vec{\tau} \cdot
\check{r} )}$.

The second and fourth terms of (\ref{deltaec}) require numerical calculations.
In
the first
case,
the sum over the index $k$ has been cutted when the tail of the series becomes
negligible.
Integrations in the $x$ variable have been numerically solved in both cases.

\bigskip

As it was said before, the reference (\mitbag ) Casimir energy for fermions and
that
corresponding
to gluons have been studied in a complementary way. In reference \cite{mitdef},
the
free energy
for a fermionic and an abelian gauge field (enough for the 1-loop description
of the
free
energy
for gluons) was studied. Such evaluations have been made using analytically
regularized
traces which involve the Green functions of the boundary problems considered.

The $T \rightarrow 0$ limit of the results in \cite{mitdef} has a simple
structure.
As it can
be understood by dimensional analysis, the Casimir energy (once singularities
have
been
removed by the renormalization of the zero-loop energy) takes the form
\beq
e_{c,MIT} = R E_{c,MIT} = K_{MIT} \, ,
\label{ecmit}
\eeq
where $K_{MIT}$ is the arbitrary finite part left by the renormalization
procedure.

We thus have now the total Casimir energy of the bag, including the correction
due
to the
interaction of fermions with the external Skyrme field, represented by the
chiral angle $\theta$,
\beq
e_c (\theta) = \Delta e_{c} (\theta) + e_{c,MIT} \, ,
\label{cas-total}
\eeq
up to the knowledge of the constants $K_Q$ and $K_{MIT}$. As we will show
later, they can be determined imposing physical conditions in the framework  of
the
two phase
chiral bag model.

\bigskip

A further contribution has to be included if the total inner bag energy (energy
of
the
defect, in the remaining of the paper) is studied.

In reference \cite{goldstone}, it has been proved that the valence quark
contribution must be
taken into account in the $ 0 \leq \theta \leq \pi/2$ region, to obtain the
baryon
number,
$\cal B$, in the
TPM. In fact, when adding the contributions coming from the truncated Skyrme
model,
the Dirac
sea and the
valence quarks, one obtains ${\cal B}=1$ for any value of $\theta$.
This can be understood by studying the energy of the fundamental eigenstate of
the
Dirac
Hamiltonian \cite{mulders}. When $\theta < \pi/2$, the fundamental energy
becomes
positive and
valence quarks leave the Dirac sea. In this case, they must be explicitely
included.
As
regards the bag energy, the
valence
quark contribution can be written as
\beq
e_q = 3 \epsilon (\theta) H(\pi/2 - \theta) \, ,
\label{valencia}
\end{equation}
where $\epsilon (\theta)$ is the fundamental eigenvalue and $H(x)$ is the
Heaviside
step
function.

\bigskip

Note that our regularization prescription lead to a finite bag energy
(depending on
$K_Q$ and
$K_{MIT}$). All derived quantities will also be finite. Such
is the case of the axial flux through the boundary of the sphere,
\beq
\phi^f (R, \theta) =
\int_{r=R} d\Omega \, n_\mu n_a \left< j_5^{\mu,a} \right> =
\frac1R
\frac{d}{d \theta} e_t (\theta) \, ,
\label{flujofermiones}
\end{equation}
where $e_t (\theta) = \left[ \Delta e_c (\theta) +
e_q (\theta) + K_{MIT} \right]$.
The last term, coming from the \mitbag \ bag, is $\theta$-independent and does
not
contribute to
the
axial flux.

As it was proposed in reference \cite{vepstas2}, the vanishing of the flux when
$R
\rightarrow
0$ determines the $K_Q$ renormalized constant. This physical imposition is
consistent with
the analysis we will perform later, when treating the TPM.

\bigskip

The external phase of the TPM is described by the Skyrme model
\cite{skyrme,witten}. The lagrangian can be written as
\beq
{\cal L} = \frac1{16} F_\pi^2 \tram \left( \partial_\mu U \partial^\mu
U^{\dag} \right) +
\frac{1}{32 e^2} \tram \left[ \left( \partial_\mu U \right) U^{\dag},
\left( \partial_\nu U
\right) U^{\dag} \right]^{2} \, ,
\label{skyrme}
\eeq
where the scalar field $U(x)$ takes its values in the $SU(2)$ group.

Two parameters have been introduced here: $F_\pi$, associated with the pion's
decay
constant
(experimental value $F_\pi^{exp} = 186 \, {\rm MeV}$) and $e^2$, which
represents
the strength of
the stabilizing term. These parameters will be adjusted later in the framework
of
the two
phase model \cite{vento}.

The Skyrmion is a topologically stable classical solution of the Lagrangian in
(\ref{skyrme}),
when the whole space is considered. It is given by

\begin{equation}
U_0 (\vec{x}) = e^{i \theta (r) (\vec{\tau} \cdot \vec{x})} \, ,
\label{ansatz}
\end{equation}
where $\theta (r)$ is what we are calling the chiral angle. Spatial and
isospinorial
indexes
are
linked in the argument of the exponential.

By the imposition of the boundary conditions
\begin{equation}
\theta (r=0) = \pi \qquad \theta (r) \rightarrow_{r \rightarrow \infty} 0 \, ,
\label{condiciones}
\end{equation}
an skyrmion of topological baryonic number (winding number) ${\cal B} = 1$ is
obtained (in the
pure Skyrme model).

Replacing the Skyrme ansatz in the Lagrangian (\ref{skyrme}), the equation of
motion
is
obtained as a nonlinear differential equation for $\theta (r)$ \cite{adkins}.

The Skyrme lagrangian is invariant under $SU(2)_L \otimes SU(2)_R$ chiral
transformations
\begin{equation}
U \rightarrow AUB^{-1} \, ,
\label{transfaxial}
\end{equation}
where $A$ and $B$ belong to $SU(2)$. When $A=B^{-1}$, we are in the case of
axial
symmetry
leading to the locally conserved axial current. Chiral boundary conditions
guarantee
its
conservation
even at the boundary, when the TPM is considered.

The flux of the axial current through an sphere of radius ${r}$, in terms of
the
adimensionalized
radius $\hat{r} = \frac{e F_\pi}{2} r$, is given by
\beq
\phi^{Sk} (\hat{r}) = \frac{2 \pi F_\pi}{e} \frac{d \theta}{d \hat{r}}
\hat{r}^2 \left[ 1 + \frac{2}{\hat{r}^2} \sin^2 \theta \right] \, .
\label{flujoskyrmion}
\eeq

It is not difficult to show that, when $R \rightarrow 0$ ($\theta \rightarrow
\pi,
\theta^{\prime} < \infty$ ) $\phi^{Sk} \rightarrow 0$. It is reasonable  to
extend
such
behavior to the flux from the inner fermionic phase, using this criterium to
determine the
value of the
renormalized constant $K_Q$.

\bigskip

In the TPM, the Skyrmion is truncated to the exterior of an sphere of radius
$R$.
In the $R \rightarrow 0$ limit, the pure skyrmion should describe the baryon
properties. The contribution $K_{MIT}/R$ to the Casimir energy is forbidden in
such
scheme.
Then,
the validity of the hybrid chiral model, even in the $R \rightarrow 0$ limit,
imposes $K_{MIT}
= 0$.

\bigskip

Once our renormalization scheme has been established, a fit can be proposed for
the
numerically evaluated
Casimir energy. Following \cite{vepstas2} we propose
\[
e_c (\theta) -
3 K_Q \sin^2 \theta = 3 \left\{
\frac{3}{4 \pi} \left( \left\{
\begin{array}{cc}
\theta^2 & 0 \leq \theta \leq \frac{\pi}2 \\
(\pi - \theta )^2 & \frac{\pi}{2} \leq \theta \leq \pi
\end{array}
\right\} -
\sin^2 \theta \right) \right.
\]
\beq
\left. + C_2 \sin^2 \theta + C_4 \sin^4 \theta +
C_6 \sin^6 \theta + C_8 \sin^8 \theta \right\} \, ,
\label{ajuste}
\eeq
where the coefficients take the values
\beq
\begin{array}{crccr}
C_2 = & -0.13381 &\qquad & C_4 = & 0.05085 \\
C_6 = & -0.01257 &\qquad & C_8 = & 0.01241
\end{array} \, .
\label{coeficientes}
\eeq

The required vanishing of the axial flux of fermions in the $R \rightarrow 0$
limit
inmediately leads to
\[
K_Q = - C_2 \, ,
\]
thus eliminating the contribution proportional to $(\pi-\theta)^2$ for
$R\rightarrow
0$.

\begin{figure}
\epsffile{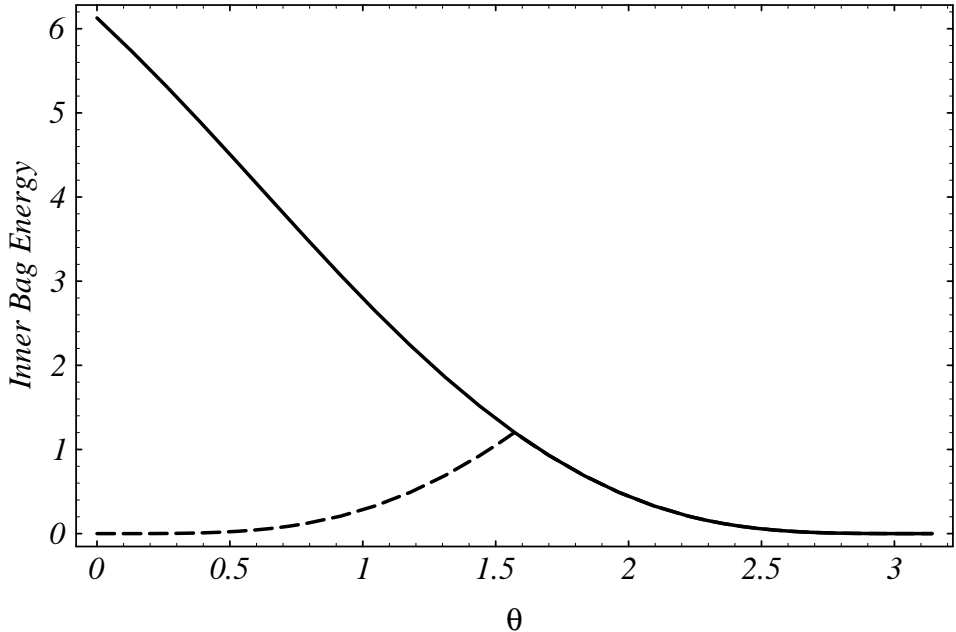}
\caption{(Dimensionless) Inner energy of the chiral bag ;
-- -- : Casimir energy ; --- : Total inner energy}
\label{fig-casimir}
\end{figure}

Having fixed the renormalized constants, the energy of the inner phase is as
shown
in Figure~\ref{fig-casimir}. Our results are totally consistent with those
presented
in
\cite{vepstas2}.
The symmetry  of the Casimir energy (dashed line) about $\theta=\pi/2$ is
evident. Such property was a priori expectable from simple properties of
the
eigenvalues of the Dirac hamiltonian of the model.

In the same figure, the solid line represents the total energy of the inner
phase
(when the
valence quark
contribution is also taken into account). As in reference \cite{vepstas2}, the
resulting interior energy is a smooth function.

\bigskip

In the TPM picture, the axial flux through the boundary of the defect should be
continuous for
all
values of the bag radius. So, a fine tunning of the strength of the
stabilization term $e(R)$ must be performed to ensure
\beq
\phi^f (R) = \phi^{Sk} (R) \qquad {\rm for \quad all \quad R} \, .
\label{igualdad-flujos}
\eeq

To impose this condition, the knowledge of the chiral angle as a function of
$R$,
$\theta
(R)$, is necessary. As it
was said,
its value can be obtained numerically, by solving a nonlinear differential
equation.
But,
following
the proposition of M.~Atiyah and N.~Manton \cite{atiyah}, $U(\vec{x})$
configurations can be
constructed by evaluating the holonomy of Yang-Mills fields with topological
charge
$k$, in the time direction. This is derived from a t'Hooft instanton of width
$\lambda$.

For the $k=1$ case, the resulting chiral angle is
\beq
\theta (\hat{r}) = \pi \left[ 1 - \frac{1}{\sqrt{1+ \left(
\frac{\hat{\lambda}}{\hat{r}}
\right)^2}} \right] \, .
\label{perfil}
\eeq

Replacing $\theta (\hat{r} )$ in the energy of the skyrmion\cite{adkins}
\[
M_{Sk} = - \int d^3x \, {\cal L}_{Sk}
\]
\beq
=
4 \pi \int_{0}^{\infty} \left\{ \frac{F_\pi^2}{8} \left( r^2
\theta^{\prime 2} + 2 \sin^2
\theta \right) + \frac{1}{2 e^2} \frac{\sin^2 \theta}{r^2} \left\{ 2 r^2
\theta^{\prime 2} +
\sin^2 \theta \right) \right\} \, ,
\label{masa}
\eeq
and minimizing with respect to $\hat{\lambda}$, the value of this parameter is
fixed
to
$\hat{\lambda}=1.452$
\cite{atiyah}. The resulting profile is very close to the one obtained
numerically.
For computational convenience, we
will
use this aproximation for the rest of the paper.

With the Atiyah-Manton profile, equation (\ref{igualdad-flujos}) gives
\beq
\frac1{e^2 (\theta )} =
- 8 \frac{3 \pi}{32 \hat{\lambda}^2} \frac{1}{(\pi -
\theta)^3
\left[ 1 + \frac2{\hat{\lambda}^2} \frac{\theta ( 2 \pi - \theta )}{(\pi
- \theta )^2}
\sin^2 \theta \right]} \frac{d e_t (\theta)}{d \theta} \, ,
\label{etheta}
\eeq
when parametrized by the chiral angle $\theta$.

When the $\theta \rightarrow \pi$ ($R \rightarrow 0$) limit is taken, the
stabilization
strength for the pure skyrmion is obtained
\beq
e(R=0) = 4.216 \, ,
\eeq
to be compared with the value $e=5.45$ obtained by G.~S.~Adkins, C.~R.~Nappi
and
E.~Witten by
fitting the masses of the nucleon and the $\Delta$ particle.

Now, the value of $F_\pi$ can be fixed in the scheme proposed in \cite{adkins}.
Having $e$ and
the nucleon mass $M_n^{exp} = 938 {\rm MeV}$, expression (9) of \cite{adkins}
(with
$\lambda$ and $M$ calculated for
the Atiyah-Manton profile) leads to $F_\pi=99.59 \, {\rm MeV}$. This value, far
from the
experimental one, is, however, near $F_\pi = 129 \, {\rm MeV}$ obtained in
\cite{adkins}. As a
consistency check, the $\Delta$ particle mass has been calculated with our
parameters, giving
$M_\Delta = 1206 \, {\rm MeV}$ (experimental value: $M_\Delta ^{exp.}= 1230 \,
{\rm
MeV}$).

\bigskip

\begin{figure}
\epsffile{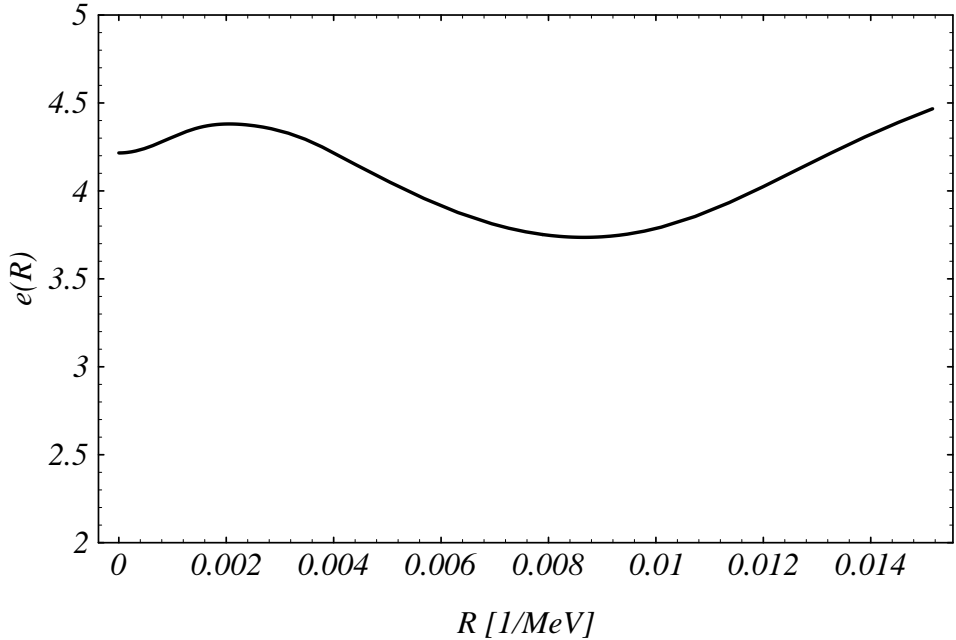}
\caption{Strength of the stabilization term $e(R)$}
\label{fig-er}
\end{figure}

The calculations just detailed complete the determination of the parameters of
the
Skyrme
model, in a
consistent way with the TPM under study.
Now, recovering the dimensional variable $r$, the strength $e(R)$ is as shown
in
Figure~\ref{fig-er}.

\bigskip

To complete our task, the energy contained in the Skyrme external sector must
be
evaluated. To do this calculation, the integral in (\ref{masa}), with the lower
limit
truncated to the radius of the defect, must be studied.
We have performed the numerical evaluation of this quantity using the
Atiyah-Manton
profile
and the
parameters of the Skyrme model $F_\pi$ and $e(R)$, thus obtaining the energy in
the
Skyrme
sector as a
function of the defect radius.

\begin{figure}
\epsffile{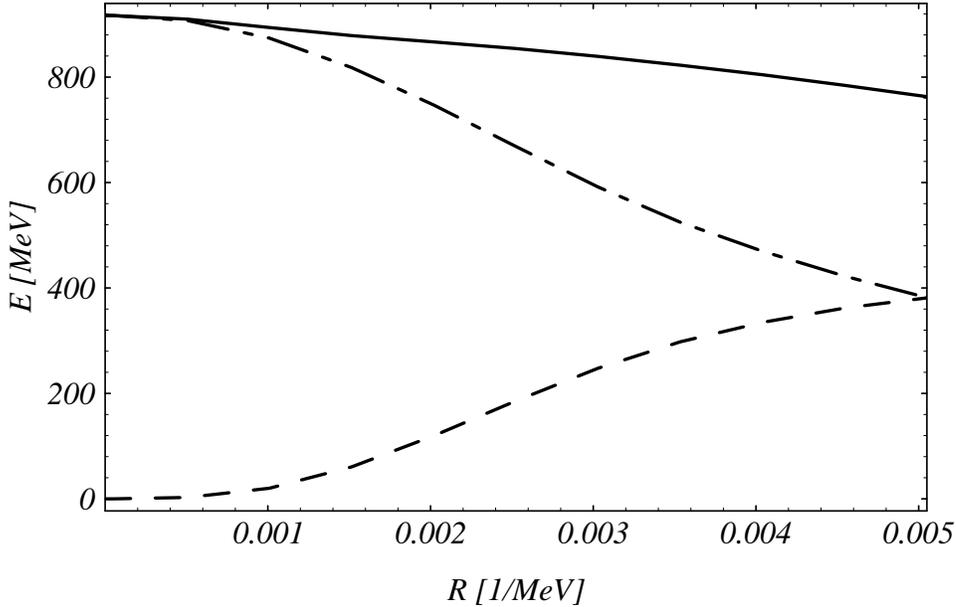}
\caption{Two Phase model energy
--- : Complete model ; -- $\cdot$ -- : Skyrmion's sector ; -- -- : Bag's
sector}
\label{fig-cheshire}
\end{figure}

Figure~\ref{fig-cheshire} is the main result of this paper. It shows in dashed
lines, from top
to bottom, the energy of the external Skyrme phase and the energy of the
defect, as
functions
of the position of the limit between the phases. Also shown, in solid line, is
the
total energy
of the hybrid
chiral bag model. This last shows a remarkable independence with the bag
radius (for $0 \leq R \leq 1 \, {\rm fm}$), as suggested by the CCP.

\bigskip

Summarizing, we have employed the results presented in
\cite{bolsa-quiral,mitdef}
for the
internal Casimir energies in a chiral bag model. In these papers, by the use of
analytical regularizations, a renormalized Casimir energy, dependent on the
$K_Q$
and $K_{MIT}$
constants, was obtained.
In the present paper, a TPM was completed, by introducing an external Skyrme
phase.
The
renormalized constants $K_Q$ and $K_{MIT}$, as well as the parameters $F_\pi$
and
$e(R)$ of the
truncated Skyrme Lagrangian were determined according to physical conditions,
suitable for the
TPM.

In reasonable agreement with the Cheshire Cat hypothesis, the total energy of
this
model shows
an
approximate independence
with the bag radius (separation limit between the phases) in the range of $0
\leq R
\leq 1 \,
{\rm fm}$.

\bigskip

The study of CBM at finite temperature \cite{deconf,loewe} is an interesting
effective
approach to the analysis of deconfinement transitions. In those references,
succesive
approximations to the problem,
based on the validity of the Cheshire Cat hypothesis at $T=0$ have been made.
The
present
results give a ground to such hypothesis, thus making it sensible to
look for the presence of deconfinement transitions only in the
temperature-dependent
contributions to the free energy of the bag.

\end{document}